\documentstyle[12pt,epsfig]{article}
\begin{document}
\baselineskip=24pt
\def\rd{{\rm d}}
\newcommand{\nn}{\nonumber}
\newcommand{\ra}{\rightarrow}
\renewcommand{\baselinestretch}{1.5}
\begin{titlepage}
\vspace{-20ex}
\vspace{1cm}
\begin{flushright}
\vspace{-3.0ex} 
    {\sf ADP-02-93/T531} \\
\vspace{-2.0mm}
\vspace{5.0ex}
\end{flushright}

\centerline{\Large\sf Chiral Extrapolation of Lattice Data for Heavy Baryons} 
\vspace{6.4ex}
\centerline{\large\sf  	X.-H. Guo and A.W. Thomas}
\vspace{3.5ex}
\centerline{\sf Department of Physics and Mathematical Physics,}
\centerline{\sf and Special Research Center for the Subatomic Structure of
Matter,}
\centerline{\sf Adelaide University, SA 5005, Australia}
\centerline{\sf e-mail:  xhguo@physics.adelaide.edu.au,
athomas@physics.adelaide.edu.au,}
\vspace{6ex}
\begin{center}
\begin{minipage}{5in}
\centerline{\large\sf 	Abstract}
\vspace{1.5ex}
\small {The masses of heavy baryons containing a $b$ quark have been 
calculated numerically in lattice QCD with pion masses 
which are much larger than its physical value. In the present work we 
extrapolate these lattice data to the physical mass of the pion 
by applying the effective chiral Lagrangian for heavy baryons, 
which is invariant under chiral symmetry when the light quark masses go to 
zero and heavy quark symmetry when the heavy quark masses go to infinity. 
A phenomenological functional form with three parameters, which has the
correct behavior in the chiral limit and appropriate behavior when the
pion mass is large, is proposed to 
extrapolate the lattice data. It is found that the extrapolation 
deviates noticably from the naive linear extrapolation  
when the pion mass is smaller than about 500MeV. The mass differences 
between $\Sigma_b$ and $\Sigma_b^*$ and between $\Sigma_b^{(*)}$
and $\Lambda_b$ are also presented. Uncertainties arising from both lattice
data and our model parameters are discussed in detail. We also give a 
comparision of the results in our model with those obtained in 
the naive linear extrapolations.}

\end{minipage}
\end{center}

\vspace{0.5cm}

\end{titlepage}

\vspace{0.2in}
{\large\bf I. Introduction}
\vspace{0.2in}

The spectrum of some hadrons has been calculated numerically in lattice
QCD over the past few years. These hadrons include light mesons and
baryons \cite{light}, heavy mesons \cite{khan1, hein}, and heavy baryons
\cite{khan1, mathur}. Using non-relativistic QCD (NRQCD) on the lattice
\cite{lepage} for heavy quarks and the tadpole-improved clover action for 
light quarks, the authors of Refs.\cite{khan1, hein} studied extensively 
the spectra of heavy mesons and heavy baryons (including doubly heavy baryons).
These lattice data were obtained in the region where the mass of the pion 
is much larger than the physical mass of the pion. Hence one needs to
extrapolate these data to the physical pion mass in order to obtain the 
heavy hadron masses in the real world. Naively, this is
done by linear extrapolations which are inconsistent with the model 
independent, non-analytic behavior of hadron 
properties in the chiral limit. In order to overcome this problem,
pion-hadron loops are included in the study of light hadron properties 
\cite{lein1, lein2, detmold, jones}. This yields the correct leading and 
next-to-leading non-analytic terms in the light quark masses and
leads to rapid variation at small pion masses. In general, lattice data  
extrapolated to the physical pion mass this way yield quite different results
from linear extrapolations. Based on this,
we considered previously the chiral extrapolation of the lattice data for 
heavy $D$ and $B$ mesons and discussed the important hyperfine 
splittings \cite{guo}. Here we generalize 
our approach to the case of heavy baryons and extrapolate the lattice data
for heavy $b$-baryons obtained in Ref.\cite{khan1}.

We work in two opposite limits of quark masses. One is the zero quark mass
limit while the other is the infinite quark mass limit. When  
the masses of the light quarks, $u$, $d$, and $s$, 
go to zero the QCD Lagrangian
has a chiral $SU(3)_L \times SU(3)_R$ symmetry, which is spontaneously broken 
into $SU(3)_V$ plus eight Goldstone bosons. When the masses of the heavy
quarks $c$ and $b$ go to infinity, we have an effective theory,
heavy quark effective theory (HQET), which is invariant under 
heavy quark flavor and heavy quark spin transformations,
$SU(2)_f \times SU(2)_s$. Thus the interactions of heavy baryons with
the light pseudoscalar mesons should be described by an effective chiral 
Lagrangian for heavy baryons which is invariant under both  
$SU(3)_L \times SU(3)_R$ and $SU(2)_f \times SU(2)_s$ transformations.
This chiral Lagrangian will be applied in the small pion mass region
while we extrapolate the lattice data to the physical pion mass.

The remainder of this paper is organized as follows. In Section II we 
give a brief review of the chiral Lagangian for heavy baryons
and including the propagators of heavy baryons.
In Section III we apply this Lagrangian to calculate pion loop contributions to
the self-energy of heavy baryons. Then we   
propose a phenomenological functional form with three parameters 
for extrapolating the lattice data to the  
physical region.  In Section IV we use this form to fit the
lattice data and give numerical results. Finally, Section V contains a 
summary and discussion.
 
\vspace{0.2in}
{\large\bf II. Chiral perturbation theory for heavy baryons}
\vspace{0.2in}

When the light quark mass, $m_q$, approaches zero, the QCD Lagrangian 
possesses an $SU(3)_L \times SU(3)_R$
chiral symmetry. The light pseudo-Goldstone bosons associated
with spontaneous breaking of chiral symmetry can be incorporated into a
$3 \times 3$ matrix 
\begin{equation}
\Sigma = {\rm exp} \left( \frac{2iM}{f_\pi} \right),
\label{2a}
\end{equation}
where $f_\pi$ is the pion decay constant, $f_\pi=132$MeV, and $M$ 
is a matrix which includes the eight Goldstones 
\begin{eqnarray}
M=\left (
\begin{array}{ccc}
\frac{1}{\sqrt{2}}\pi^0+\frac{1}{\sqrt{6}}\eta & \pi^+ & K^+ \\
\pi^- &-\frac{1}{\sqrt{2}}\pi^0+\frac{1}{\sqrt{6}}\eta &K^0\\
K^- &\bar{K}^0 &-\sqrt{\frac{2}{3}}\eta\\
\end{array}
\right).     
\label{2b}
\end{eqnarray}
Under $SU(3)_L \times SU(3)_R$ transformations, $\Sigma$ is required
to transform linearly,
\begin{equation}
\Sigma \ra L \Sigma R^+,
\label{2c}
\end{equation}
where $L \in SU(3)_L$ and $R \in SU(3)_R$. 

While discussing the interactions of Goldstone bosons with other matter
fields it is convenient to introduce 
\begin{equation}
\xi=\sqrt{\Sigma}.
\label{2d}
\end{equation}
Under $SU(3)_L \times SU(3)_R$ transformations
\begin{equation}
\xi \ra L \xi U^+=U\xi R^+,
\label{2e}
\end{equation}
where the unitary matrix $U$ is a complicated nonlinear function of $L$, $R$, 
and the Goldstone fields, and is invariant under the parity transformation.

A heavy baryon is composed of a heavy quark $Q$ ($Q=b$, or $c$) and 
two light quarks $q_a q_b$ ($a$ ($b$) equals 1, 2, 3 for $u$, $d$, $s$ 
quarks, respectively). When the heavy quark mass, $m_Q$, is much larger 
than the QCD scale, $\Lambda_{\rm QCD}$, the light degrees of freedom in a 
heavy baryon become blind to the flavor and spin quantum
numbers of the heavy quark because of the $SU(2)_f \times 
SU(2)_s$ symmetries. Therefore, the light degrees of freedom have good quantum 
numbers which can be used to classify heavy baryons. 
The angular momentum and parity $J^P$ of the two light quarks may be 
$0^+$ or $1^+$, which correspond to $SU(3)_{L+R}$ antitriplet and 
sextet, respectively. The lowest-lying heavy baryons in the $\bar{3}$
representation have spin 1/2, and are denoted by fields which destroy these
baryons, $T_a$ ($T_3=\Lambda_Q$, $T_{1,2}=\Xi_{Q}^{\prime}$). 
The lowest-lying heavy baryons in the 6 representation have spin 1/2
or 3/2, and are denoted by field operators $S^{a b}$ and $S_{\mu}^{a b}$,
respectively, where $S_{(\mu)}^{(*) 11, 12, 22}=\Sigma_{Q}^{(*)}$, 
$S_{(\mu)}^{(*) 13, 23}=\Xi_{Q}^{(*)}$, and  
$S_{(\mu)}^{(*) 33}=\Omega_{Q}^{(*)}$.
$T_a$ transforms under $SU(3)_L \times SU(3)_R$ as
\begin{equation}
T_a \ra T_b U^{+}_{ba},
\label{2f}
\end{equation}
and under heavy quark spin symmetry
\begin{equation}
T_a \ra S T_a,
\label{2g}
\end{equation}
where $S \in SU(2)_s$. $T_a$ also satisfies 
\begin{equation}
T_a =\rlap/v T_a,
\label{2h}
\end{equation}
where $v$ is the velocity of the heavy baryon.

It is convenient to combine $S^{a b}$ and $S_{\mu}^{* a b}$ into the field
$S_{\mu}^{a b}$
\cite{georgi}
\begin{equation}
S_{\mu}^{a b}=\frac{1}{\sqrt{3}}(\gamma_\mu +v_\mu)\gamma_5 S^{a b}+
S_{\mu}^{* a b}.
\label{2i}
\end{equation}
Then under $SU(3)_L \times SU(3)_R$ 
\begin{equation}
S_{\mu}^{a b} \ra U^a_c U^b_d S_{\mu}^{c d},
\label{2j}
\end{equation}
while under heavy quark spin symmetry
\begin{equation}
S_{\mu}^{a b} \ra S S_{\mu}^{a b}.
\label{2k}
\end{equation}
$S_{\mu}^{a b}$ satisfies the constraints
\begin{equation}
S_{\mu}^{a b} =\rlap/v S_{\mu}^{a b},\;\; v^\mu S_{\mu}^{a b}=0.
\label{2h2}
\end{equation}
 
It is convenient to introduce a vector field $V_{ab}^\mu$,
\begin{equation}
V_{ab}^\mu=\frac{1}{2}(\xi^+ \partial^\mu \xi +\xi \partial^\mu \xi^+)_{ab},
\label{2l}
\end{equation}
and an axial-vector field $A_{ab}^\mu$,
\begin{equation}
A_{ab}^\mu=\frac{i}{2}(\xi^+ \partial^\mu \xi -\xi \partial^\mu \xi^+)_{ab}.
\label{2m}
\end{equation}
Under $SU(3)_L \times SU(3)_R$, $V^\mu \ra U V^\mu U^+ +U \partial^\mu U^+$,
and $A^\mu \ra U A^\mu U^+$. Defining the covariant derivative
\begin{equation}
(D^\mu T)_a=\partial^\mu T_a -T_b (V^\mu)^b_a,
\label{2n}
\end{equation}
and 
\begin{equation}
(D^\mu S_\nu)^{a b}=\partial^\mu S_\nu^{a b} +(V^\mu)^a_c S_\nu^{c b}
+(V^\mu)^b_c S_\nu^{a c},
\label{2o}
\end{equation}
we can show that under $SU(3)_L \times SU(3)_R$, $D^\mu T_a \ra D^\mu T_b 
(U^+)^b_a$ and $D^\mu S_\nu^{a b} \ra U^a_c U^b_d D^\mu S_\nu^{c d}$.
 
In the limit where light quarks have zero mass and heavy quarks have 
infinite mass, the Lagragian for the strong interactions of heavy baryons 
with Goldstone pseudoscalar
bosons should be invariant under both chiral symmetry and heavy quark
symmetry. It should also be
invariant under Lorentz and parity transformations as required in general.
The most general form for the Lagragian satisfying these requirements is
\cite{cho}
\begin{eqnarray}
{\cal L}&=& i \bar{T}^a v_\nu (D^\nu T)_a -i\bar{S}^\mu_{a b} v_\nu 
(D^\nu S_\mu)^{a b} + \Delta M \bar{S}^\mu_{a b} S_\mu^{a b}
+i g_1 \epsilon_{\mu\nu\sigma\lambda}\bar{S}^\mu_{a c}v^\nu 
(A^\sigma)^a_b S^{\lambda b c}\nn\\
&&+g_2[\epsilon_{a b c}\bar{T}^a (A^\mu)^b_d S_\mu^{c d}+
\epsilon^{a b c}\bar{S}^\mu_{c d} (A_\mu)^d_b T_a],
\label{2p}
\end{eqnarray}
where $g_1$ and $g_2$ are coupling constants describing the interactions 
between heavy baryons and Goldstone bosons and $\Delta M$ is the mass
difference between sextet and antitriplet heavy baryons in the heavy
quark limit. As a consequence of heavy quark symmetry, 
$g_1$ and $g_2$ are universal for different heavy baryons.
Since they contain information about the
interactions at the quark and gluon level, they cannot be fixed from chiral
perturbation theory, but should be determined by experiments.  

In the limit $m_Q \ra \infty$, the propagator for $\Lambda_Q$ is 
$$\frac{i}{v \cdot p} \frac{1+\rlap/v}{2},$$ 
where $p$ is the residual momentum of the heavy baryon. 
It can also be shown that for $\Sigma_Q$, the propagator is
$$\frac{i}{v \cdot p-\Delta M} \frac{1+\rlap/v}{2},$$ and  
the propagator for $\Sigma_Q^*$ is
$$\frac{1+\rlap/v}{2}\frac{-i(g_{\mu\nu}-\frac{1}{3}\gamma_\mu \gamma_\nu
-\frac{2}{3}v_\mu v_\nu)}{v \cdot p-\Delta M}\frac{1+\rlap/v}{2}.$$ 
In the limit $m_Q \ra \infty$, there is no mass difference between 
$\Sigma_Q$ and $\Sigma_Q^*$. 

In HQET, the leading term which is responsible for a mass difference between
$\Sigma_Q$ and $\Sigma_Q^*$ is the color-magnetic-moment operator, 
$\frac{1}{m_Q} \bar{h}_v \sigma_{\mu\nu}G^{\mu\nu} h_v$ 
(where $h_v$ is the heavy quark 
field operator in HQET and $G^{\mu\nu}$ is the gluon field strength tensor).
This term is singlet under $SU(3)_L \times SU(3)_R$ and leads to 
the following correction term to ${\cal L}$ in Eq.(\ref{2p}):
\begin{equation}
i\frac{\alpha}{m_Q} \bar{S}^\mu_{a b}\sigma_{\mu \nu} S^{\nu a b},
\label{2q}
\end{equation}
where $\alpha$ is a constant which also contains interaction information
at the quark and gluon level, and which is the same for 
$\Sigma_Q$ and $\Sigma_Q^*$ at the tree level because of heavy quark symmetry. 
When QCD loop corrections are included, $\alpha$ 
depends on $m_Q$ logarithmically.

The term (\ref{2q}) enhances the mass of $\Sigma_Q^*$ by $\alpha/m_Q$ and
lowers that of $\Sigma_Q$ by $2\alpha/m_Q$. Therefore, the propagators
for $\Sigma_Q$ and $\Sigma_Q^*$ become
$$\frac{i}{v \cdot p-\Delta M+\frac{2\alpha}{m_Q}} \frac{1+\rlap/v}{2},$$ and
$$\frac{1+\rlap/v}{2}\frac{-i(g_{\mu\nu}-\frac{1}{3}\gamma_\mu \gamma_\nu
-\frac{2}{3}v_\mu v_\nu)}{v \cdot p-\Delta M -\frac{\alpha}{m_Q}}
\frac{1+\rlap/v}{2},$$
respectively.

Substituting $\xi={\rm exp}(iM/f_\pi)$ into Eqs.(\ref{2l}, \ref{2m}) and
making a Taylor expansion we obtain the following expressions for  $V_\mu$ and 
$A_\mu$:
\begin{equation}
V_\mu=\frac{1}{2f_\pi^2}[M, \partial_\mu M] + O(M^4),
\label{2r}
\end{equation}
\begin{equation}
A_\mu=-\frac{1}{f_\pi} \partial_\mu M +O(M^3).
\label{2s}
\end{equation}

Substituting Eq.(\ref{2i}) and  Eqs.(\ref{2r}, \ref{2s}) 
into Eq.(\ref{2p}) we have the
following explicit form for the interactions of heavy baryons with Goldstone
bosons:
\begin{eqnarray}
&&-\frac{i}{2f_\pi^2} \bar{T}^a [M, v \cdot \partial M]_a^b T_b+
\frac{i}{2f_\pi^2} \left\{\bar{S}_{a b} [M, v \cdot \partial M]^a_c S^{c b}
+\bar{S}_{a b} [M, v \cdot \partial M]^b_c S^{a c}
 \right.\nn\\
&&\left. -\bar{S}^{* \mu}_{a b} [M, v \cdot \partial M]^a_c S^{* c b}_{\mu}
-\bar{S}^{* \mu}_{a b} [M, v \cdot \partial M]^b_c S^{* a c}_{\mu}
\right\}-\frac{i}{f_\pi}g_1 \epsilon_{\mu\nu\sigma\lambda}v^\nu 
(\partial^\sigma M)^a_b  \nn\\
&&\times \left[-\frac{1}{3} \bar{S}_{a c} \gamma^\mu
\gamma^\lambda S^{b c}+\bar{S}_{a c}^{* \mu} S^{* \lambda b c}
+\frac{1}{\sqrt{3}} \bar{S}^{* \mu}_{a c}\gamma^\lambda \gamma_5
S^{b c}- \frac{1}{\sqrt{3}} \bar{S}_{a c}\gamma_5 \gamma^\mu
S^{* \lambda b c}\right] \nn\\
&&-\frac{g_2}{f_\pi}\left[\frac{1}{\sqrt{3}}
\epsilon_{a b c}
\bar{T}^a (\partial^\mu M)^b_d (\gamma_\mu +v_\mu)\gamma_5 S^{c d}
+\epsilon_{a b c}\bar{T}^a (\partial^\mu M)^b_d S^{* c d}
 \right.\nn\\
&&\left.-\frac{1}{\sqrt{3}} \epsilon_{a b c} \bar{S}^{c d}\gamma_5
(\gamma^\mu +v^\mu)(\partial_\mu M)^b_d T^a+\epsilon_{a b c} 
\bar{S}^{* \mu c d} (\partial_\mu M)^b_d T^a\right],
\label{2t}
\end{eqnarray}
where $O(M^3)$ terms are ignored.

Chiral symmetry can be broken explicitly by nonzero light quark masses.
This leads to the following leading order terms in the
explicit chiral symmetry breaking masses:
\begin{eqnarray}
&&\lambda_1 \bar{S}^{\mu}_{a b} (\xi m_q \xi +\xi^+ m_q \xi^+)^b_c 
S^{a c}_\mu +\lambda_2 \bar{S}^{\mu}_{a b} S^{a b}_\mu 
{\rm Tr}(m_q \Sigma^+ + \Sigma m_q)\nn\\
&& +\lambda_3 {\rm Tr} T_a (\xi m_q \xi +\xi^+ m_q \xi^+)^a_b \bar{T}^b
+\lambda_4 {\rm Tr} (\bar{T}^a T_a){\rm Tr}(m_q \Sigma^+ +\Sigma m_q),
\label{2u}
\end{eqnarray}
where $\lambda_i$ ($i=1,2,3,4$) are parameters which are also
independent of the heavy quark mass in the limit $m_Q \ra \infty$.

\vspace{0.2in}
{\large\bf III. Formalism for the extrapolation of lattice data for
heavy baryon masses}
\vspace{0.2in}

From the chiral Lagrangian for the interactions of heavy
baryons with light Goldstone bosons, Eq.(\ref{2t}), we can calculate pion
loop contributions to the heavy baryon propagators near the chiral
limit - i.e., when the pion mass is not far from the chiral limit. 
This leads to a dependence of the heavy baryon masses on the pion mass. We
will concentrate on $\Sigma_Q$ and $\Sigma_Q^*$, but other heavy baryons
can be treated in the same way.

From Eq.(\ref{2t}) we can see that there are four diagrams
for pion loop corrections to the propagator of either $\Sigma_Q$ or
$\Sigma_Q^*$, and three diagrams for $\Lambda_Q$. These diagrams are shown
in Fig. 1 for $\Sigma_Q$, in Fig. 2 for $\Sigma_Q^*$,  and in Fig. 3 for
$\Lambda_Q$. It can be easily seen that Fig. 1(a), Fig. 2(a) and Fig. 3(a) 
do not contribute (this is because the integrand is of the form $k^\mu
f(k^2)$ where $k$ is the momentum of the pion in the loop, and
$f(k^2)$ is a function of $k^2$) and we will not consider them further.

Fig. 1(b) arises from the $\Sigma_Q \pi \Sigma_Q$ vertex. 
In momentum space it can be expressed as
\begin{equation}
\frac{i}{v \cdot p -\Delta M +\frac{2\alpha}{m_Q}}
(-i\Sigma_1)\frac{i}{v \cdot p -\Delta M +\frac{2\alpha}{m_Q}}
\frac{1+\rlap/v}{2},
\label{3a}
\end{equation}
where $p$ is the residual momentum of the heavy baryon $\Sigma_Q$. From 
Eq.(\ref{2t}), Fig. 1(b) takes the following form:
\begin{eqnarray}
&&-\frac{g_1^2}{18 f_\pi^2}\epsilon_{\mu\nu\sigma\lambda}
\epsilon_{\mu'\nu'\sigma'\lambda'}v^{\nu}v^{\nu'}\left(\frac{1+\rlap/v}{2}
\gamma^\mu\gamma^\lambda \frac{1+\rlap/v}{2}\gamma^{\mu'}\gamma^{\lambda'} 
\frac{1+\rlap/v}{2}\right)\nn\\
&& \left(\frac{i}{v \cdot p -\Delta M +\frac{2\alpha}{m_Q}}\right)^2
\int \frac{{\rm d}^4 k}{(2\pi)^4}
\frac{k^\sigma k^{\sigma'}}{[v \cdot (p-k)-\Delta M +\frac{2\alpha}{m_Q}]
(k^2-m_\pi^2)},
\label{3b}
\end{eqnarray}
where again $k$ is the momentum of the pion in the loop, and $m_\pi$ is the
pion mass. 

As discussed in Ref.\cite{guo}, the integral 
\begin{equation}
X^{\mu\nu} \equiv \int \frac{{\rm d}^4 k}{(2\pi)^4}
\frac{k^\mu k^\nu}{[v \cdot k - \delta](k^2-m_\pi^2)},
\label{3c}
\end{equation}
where $\delta$ is some constant, can be written as
\begin{equation}
X^{\mu\nu} = X_1 (\delta) g^{\mu\nu}+ X_2 (\delta) v^\mu v^\nu,
\label{3d}
\end{equation}
where $X_1$ and $X_2$ are Lorentz scalars, which are functions of 
$\delta$. Obviously, only the $X_1$ term 
contributes in Eq.(\ref{3b}). In the evaluation of $X_1$, 
the integration over $k_0$
was made first by choosing the appropriate contour. Then a cutoff $\Lambda$,
which characterizes the finite size of the source of the pion,
was introduced in the three dimensional integration since pion
loop contributions are suppressed when the Compton
wavelength of the pion is smaller than the source of the pion.
Since the leading non-analytic contribution of these loops
is associated with the infrared behavior of the integral, it does not
depend on the details of the cutoff. In this way, $X_1 (\delta)$ has the
following expression \cite{guo}:
\begin{eqnarray}
X_1 (\delta) &=&\frac{i}{72\pi^2}\left\{12 (m_\pi^2-\delta^2)^{3/2}\left[
{\rm arctg}\frac{\Lambda+\sqrt{\Lambda^2+m_\pi^2}-\delta}{\sqrt{m_\pi^2
-\delta^2}}-{\rm arctg}\frac{m_\pi-\delta}{\sqrt{m_\pi^2
-\delta^2}}\right] \right. \nn\\
&&\left. +3\delta(2\delta^2-3 m_\pi^2) {\rm ln}
\frac{\Lambda+\sqrt{\Lambda^2+m_\pi^2}}{m_\pi}+3\delta\Lambda\sqrt{\Lambda^2
+m_\pi^2}
+6(\delta^2-m_\pi^2) \Lambda +2 \Lambda^3 \right\}, \nn\\
&&
\label{3e}
\end{eqnarray}
when $m_\pi^2 \geq \delta^2$;
\begin{eqnarray}
X_1 (\delta) &=&\frac{i}{72\pi^2}\left\{6 (\delta^2 - m_\pi^2)^{3/2} {\rm ln}
\left(\frac{\Lambda+\sqrt{\Lambda^2+m_\pi^2}-\delta -\sqrt{\delta^2 - m_\pi^2}}
{\Lambda+\sqrt{\Lambda^2+m_\pi^2}-\delta +\sqrt{\delta^2 - m_\pi^2}}
\left|\frac{m_\pi -\delta +\sqrt{\delta^2 -m_\pi^2}}
{m_\pi -\delta -\sqrt{\delta^2 -m_\pi^2}}\right|\right) \right.\nn\\
&&\left. +3\delta(2\delta^2-3 m_\pi^2) {\rm ln}
\frac{\Lambda+\sqrt{\Lambda^2+m_\pi^2}}{m_\pi}+3\delta\Lambda\sqrt{\Lambda^2
+m_\pi^2}
+6(\delta^2-m_\pi^2) \Lambda +2 \Lambda^3 \right\},
\label{3f}
\end{eqnarray}
when $m_\pi^2 \leq \delta^2$.
In the case where $\delta=0$, 
\begin{equation}
X_1 =\frac{i}{36\pi^2}\left(3m_\pi^3 {\rm arctg}\frac{\Lambda}{m_\pi}
-3 m_\pi^2 \Lambda +\Lambda^3 \right).
\label{3g}
\end{equation}

From Eqs.(\ref{3a}) and (\ref{3b}) we have 
\begin{equation}
\Sigma_1=i\frac{2g_1^2}{3 f_\pi^2} X_1 (\Delta_1),
\label{3h}
\end{equation}
where $\Delta_1=v\cdot p -\Delta M +\frac{2\alpha}{m_Q}$.

Figs. 1(c), (d) have the same expression as in Eq.(\ref{3a}), except for 
$\Sigma_1$ being replaced by $\Sigma_2$ and $\Sigma_3$, respectively. 
In the same way, we have 
\begin{equation}
\Sigma_2=i\frac{g_1^2}{3 f_\pi^2} X_1 (\Delta_2),
\label{3i}
\end{equation}
where $\Delta_2=v\cdot p -\Delta M -\frac{\alpha}{m_Q}$.
  
Fig. 1(d) arises from the $\Sigma_Q \pi \Lambda_Q$ vertex. In this paper we
only consider $\Sigma_b^{(*)}$ since lattice data are available for them.
For $\Sigma_b^{\pm}$, $\pi^{\pm}$ appears in the pion loop, then we have
\begin{equation}
\Sigma_3^{\Sigma_b^\pm}=i\frac{g_2^2}{f_\pi^2} X_1 (\Delta_3),
\label{3j}
\end{equation}
where $\Delta_3=v\cdot p$.  For $\Sigma_b^{0}$, $\pi^{0}$ appears in the
pion loop, and we have
\begin{equation}
\Sigma_3^{\Sigma_b^0}=i\frac{g_2^2}{2 f_\pi^2} X_1 (\Delta_3).
\label{3k}
\end{equation}

Defining $\Sigma$ as the sum of $\Sigma_1$, $\Sigma_2$, and $\Sigma_3$, the
propagator of $\Sigma_b$ becomes
\begin{equation}
\frac{i}{v \cdot p-\Delta M+\frac{2\alpha}{m_Q}-\Sigma} \frac{1+\rlap/v}{2},
\label{3l}
\end{equation}
where
\begin{equation}
\Sigma^{\Sigma_b^\pm}=i\frac{2g_1^2}{3f_\pi^2} X_1 (\Delta_1)
+i\frac{g_1^2}{3f_\pi^2} X_1 (\Delta_2)
+i\frac{g_2^2}{f_\pi^2} X_1 (\Delta_3),
\label{3m}
\end{equation}
and 
\begin{equation}
\Sigma^{\Sigma_b^0}=i\frac{2g_1^2}{3f_\pi^2} X_1 (\Delta_1)
+i\frac{g_1^2}{3f_\pi^2} X_1 (\Delta_2)
+i\frac{g_2^2}{2f_\pi^2} X_1 (\Delta_3).
\label{3n}
\end{equation}

Pion loop contributions to the propagator of $\Sigma_b^*$ can be calculated
in the same way. Fig. 2(b), (c), (d) can be expressed as 
\begin{equation}
\frac{-i}{v \cdot p-\Delta M-\frac{\alpha}{m_Q}}i\Pi_i 
\frac{-i}{v \cdot p-\Delta M-\frac{\alpha}{m_Q}}
\frac{1+\rlap/v}{2}\left(g_{\mu\nu}-\frac{1}{3}\gamma_\mu \gamma_\nu
-\frac{2}{3}v_\mu v_\nu\right)\frac{1+\rlap/v}{2},
\label{3o}
\end{equation}
where $i=1,2,3$ for Fig. 2(b), (c), and (d), respectively. After some tedious
derivations, we obtain
\begin{equation}
\Pi_1=i\frac{5g_1^2}{6f_\pi^2} X_1 (\Delta_2),
\label{3p}
\end{equation}
\begin{equation}
\Pi_2=-i\frac{g_1^2}{6f_\pi^2} X_1 (\Delta_1),
\label{3q}
\end{equation}
and
\begin{equation}
\Pi_3^{\Sigma_b^{*\pm}}=i\frac{g_2^2}{f_\pi^2} X_1 (\Delta_3),
\label{3r}
\end{equation}
\begin{equation}
\Pi_3^{\Sigma_b^{*0}}=i\frac{g_2^2}{2f_\pi^2} X_1 (\Delta_3).
\label{3s}
\end{equation}
  
Define $\Pi$ as the sum of $\Pi_1$, $\Pi_2$, and $\Pi_3$, the
propagator of $\Sigma_b^*$ becomes
\begin{equation}
\frac{1+\rlap/v}{2}\frac{-i(g_{\mu\nu}-\frac{1}{3}\gamma_\mu \gamma_\nu
-\frac{2}{3}v_\mu v_\nu)}{v \cdot p-\Delta M-\frac{\alpha}{m_Q}-\Pi}
\frac{1+\rlap/v}{2},
\label{3t}
\end{equation}
where
\begin{equation}
\Pi^{\Sigma_b^{*\pm}}=i\frac{5g_1^2}{6f_\pi^2} X_1 (\Delta_2)
-i\frac{g_1^2}{6f_\pi^2} X_1 (\Delta_1)
+i\frac{g_2^2}{f_\pi^2} X_1 (\Delta_3),
\label{3u}
\end{equation}
and 
\begin{equation}
\Pi^{\Sigma_b^{*0}}=i\frac{5g_1^2}{6f_\pi^2} X_1 (\Delta_2)
-i\frac{g_1^2}{6f_\pi^2} X_1 (\Delta_1)
+i\frac{g_2^2}{2f_\pi^2} X_1 (\Delta_3).
\label{3v}
\end{equation}

In the same way, we can calculate pion loop contributions to the 
propagator of $\Lambda_b$. Fig. 3(b) and (c) can be expressed as 
\begin{equation}
\frac{i}{v \cdot p}(-i K_i) \frac{i}{v \cdot p},
\label{3w}
\end{equation}
where $i=1,2$ for Fig. 3(b) and (c), respectively. We obtain
\begin{equation}
K_1=i\frac{3g_2^2}{f_\pi^2} X_1(\Delta_1),
\label{3x1}
\end{equation}
and 
\begin{equation}
K_2=i\frac{6g_2^2}{f_\pi^2} X_1(\Delta_2).
\label{3x2}
\end{equation}

If we define $K=K_1+K_2$, then the propagator of $\Lambda_b$ becomes
\begin{equation}
\frac{i}{v \cdot p-K} \frac{1+\rlap/v}{2},
\label{3y}
\end{equation}
where
\begin{equation}
K=i\frac{3g_2^2}{f_\pi^2} X_1(\Delta_1)+i\frac{6g_2^2}{f_\pi^2} X_1(\Delta_2).
\label{3z}
\end{equation}

After the correction from $\Sigma$  is added, the propagator of $\Sigma_b$ 
is proportional to
\begin{equation}
\frac{1}{v \cdot p -m_0 - \Sigma(v \cdot p)},
\label{3aa}
\end{equation}
where $m_0$ is the mass term without $\Sigma$ correction. 

The physical mass of $\Sigma_b$, $m$, is defined by
\begin{equation}
[v \cdot p -m_0 - \Sigma(v \cdot p)]|_{v\cdot p =m}=0.
\label{3bb}
\end{equation}
Therefore, to order $O(g_1^2, g_2^2)$ we have
\begin{equation}
m= m_0 + \Sigma(v \cdot p =m_0).
\label{3cc}
\end{equation}

For $\Sigma_b^*$ and $\Lambda_b$, $\Sigma$ is replaced by $\Pi$ and $K$, 
respectively, in Eqs.(\ref{3aa}-\ref{3cc}). For $\Sigma_b$, $\Sigma_b^*$, and
$\Lambda_b$, $m_0$ is $\Delta M -\frac{2\alpha}{m_b}, \Delta M 
+\frac{\alpha}{m_b}$, and 0, respectively. Consequently, the 
pion loop contribution to the mass of $\Sigma_b$, $\sigma_{\Sigma_b}$,
has the following expression:
\begin{equation}
\sigma_{\Sigma_b^\pm}=i\frac{2g_1^2}{3f_\pi^2} X_1 (0)
+i\frac{g_1^2}{3f_\pi^2} X_1 (-\frac{3\alpha}{m_b})
+i\frac{g_2^2}{f_\pi^2} X_1 (\Delta M -\frac{2\alpha}{m_b}),
\label{3dd}
\end{equation}
and 
\begin{equation}
\sigma_{\Sigma_b^0}=i\frac{2g_1^2}{3f_\pi^2} X_1 (0)
+i\frac{g_1^2}{3f_\pi^2} X_1 (-\frac{3\alpha}{m_b})
+i\frac{g_2^2}{2f_\pi^2} X_1 (\Delta M -\frac{2\alpha}{m_b}).
\label{3ee}
\end{equation}

In the same way, 
\begin{equation}
\sigma_{\Sigma_b^{*\pm}}=i\frac{5g_1^2}{6f_\pi^2} X_1 (0)
-i\frac{g_1^2}{6f_\pi^2} X_1 (\frac{3\alpha}{m_b})
+i\frac{g_2^2}{f_\pi^2} X_1 (\Delta M +\frac{\alpha}{m_b}),
\label{3ff}
\end{equation}
and 
\begin{equation}
\sigma_{\Sigma_b^{*0}}=i\frac{5g_1^2}{6f_\pi^2} X_1 (0)
-i\frac{g_1^2}{6f_\pi^2} X_1 (\frac{3\alpha}{m_b})
+i\frac{g_2^2}{2f_\pi^2} X_1 (\Delta M +\frac{\alpha}{m_b}).
\label{3gg}
\end{equation}
 
For $\Lambda_b$, we have
\begin{equation}
\sigma_{\Lambda_b}=i\frac{3g_2^2}{f_\pi^2} 
X_1 (-\Delta M +\frac{2\alpha}{m_b})+i\frac{6g_2^2}{f_\pi^2} 
X_1 (-\Delta M -\frac{\alpha}{m_b}).
\label{3hh}
\end{equation}
In Eqs.(\ref{3dd}-\ref{3hh}), $X_1$ is given by Eqs.(\ref{3e}-\ref{3g}).

In order to extrapolate the lattice data from large $m_\pi$ to the physical
value of the pion mass, we follow the arguments proposed in Ref.\cite{guo}
where we dealt with heavy mesons. These arguments can be generalized to
the case of heavy baryons straightforwardly.
Eqs.(\ref{3dd}-\ref{3hh}) are valid when $m_\pi$ is not
far away from the chiral limit - i.e., when $m_\pi \le \Lambda$. 
As pointed out in Refs.\cite{lein1, lein2, detmold, jones, guo}, 
pion loop contributions vanish in the limit
$m_\pi \ra \infty$, and the heavy baryon mass becomes proportional to 
$m_\pi^2$ when $m_\pi$ becomes large. This behaviour is consistent with lattice
simulations. Following Refs.\cite{lein1, lein2, detmold, jones, guo}, 
we propose the following 
phenomenological, functional form for the extrapolation of lattice
data for heavy baryons:
\begin{equation}
m_{B}=a_{B} + b_{B} m_\pi^2 +\sigma_{B},
\label{3ii}
\end{equation}
for $B=\Sigma_b$, $\Sigma_b^*$ or $\Lambda_b$.

The advantage of fitting the lattice data in this way is that we can 
guarantee that our formalism has both the
correct chiral limit behavior and the appropriate 
behavior when $m_\pi$ is large, with only three parameters ($a$, $b$, and
$\Lambda$) to be determined in the fit.

Chiral symmetry is explicitly broken by the terms in Eq.(\ref{2u}).
Substituting Eqs.(\ref{2a}, \ref{2d}, \ref{2i}) into
Eq.(\ref{2u}) we have the following explicit expression:
\begin{eqnarray}
&&2\lambda_1 \sum_{a, b=1}^{3}[m_{q^a}(-\bar{S}_{a b} S^{a b}
+\bar{S}^{* \mu}_{a b}S_{\mu}^{* a b})]+2 \lambda_2 \sum_{a=1}^{3}m_{q^a}
\sum_{a, b=1}^{3}(-\bar{S}_{a b} S^{a b}
+\bar{S}^{* \mu}_{a b}S_{\mu}^{* a b})\nn\\
&&+2 \lambda_3 \sum_{a=1}^{3}m_{q^a}\bar{T}_a T^a +2 \lambda_4 
\sum_{a=1}^{3}m_{q^a} \sum_{a=1}^{3}\bar{T}_a T^a,
\label{3ll}
\end{eqnarray}
where we have made a Taylor expansion for $\xi$ and omitted $O(1/f_{\pi}^2)$
terms. It can be seen that Eq.(\ref{3ll}) does not contribute to the
mass difference between $\Sigma_Q$ and $\Sigma_Q^*$ to order $m_q$. Corrections
to this statement are of order $m_q O(1/f_{\pi}^2)$, with extra
suppression from $m_q$ with respect to the pion loop effects. They will 
therefore be
ignored. Eq.(\ref{3ll}) may contribute to the mass different between
$\Sigma_Q^{(*)}$ and $\Lambda_Q$. Such effects will be considered to be 
effectively included in the parameter $\Delta M$ in Eq.(\ref{2p}).

\vspace{0.2in}
{\large\bf IV. Extrapolation of lattice data for heavy baryon masses}
\vspace{0.2in}

The masses of $\Sigma_b$, $\Sigma_b^*$, and $\Lambda_b$ were calculated 
with the aid of NRQCD in quenched approximation in Ref.\cite{khan1}. 
Since the mass of the heavy quark
is much larger than $\Lambda_{\rm QCD}$, it becomes
an irrelevant scale for the dynamics inside a heavy hadron and is removed from
NRQCD. This makes it possible to simulate heavy baryons when 
the lattice spacing is larger than the Compton wavelength of the heavy quark.
The lattice space used is $1/a=1.92$ GeV.
For light quarks the tadpole-improved clover action was used which has 
discretization errors of order $\alpha_s a$. The value of $\beta$ which is
related to  
the bare gauge coupling is 6.0 and the lattice size is
$16^3 \times 48$. In the simulations, three different values for the
hopping parameter $\kappa$, 0.1369, 0.1375, and 0.13808, were used.
The light quark mass is related to $\kappa$ through the definition
$m_q=\frac{1}{2a}(1/\kappa -1/\kappa_c)$, with $\kappa_c=0.13917$. These
three hopping parameters correspond to three values of $m_\pi^2$:
0.6598GeV$^2$, 0.4833GeV$^2$, and 0.3141GeV$^2$, repectively. 

The heavy baryon masses were
calculated for five different values of $a M^0$ ($M^0$ is the bare
heavy quark mass): 1.6, 2.0, 2.7, 4.0, 7.0, and 10.0, where the data for the
last two values are less reliable because of large discretization errors
\cite{khan1}. The best estimate for $a M_b^0$, 2.31, was obtained 
by matching the lattice data to the mass of the $B$ meson. 
Consequently, in our fit we first extrapolate the lattice data for
$a M^0$=1.6, 2.0, 2.7, 4.0 to $a M_b^0=2.31$. This can be done by linear
extrapolation with respect to $1/M^0$ with the form $c+\frac{d}{M^0}$,
where $c$ and $d$ are constants. This is because
$a E_{\rm sim}$, which is the simulation mass in NRQCD and which is
related to the heavy baryon mass, 
depends on $1/M^0$ linearly (note that in the case of $b$-baryons,
$O((1/M^0)^2)$ can be safely ignored). Then from the data in Table XV of
Ref.\cite{khan1}, we obtain the values of $a E_{\rm sim}$ for
the three hopping parameters at $a M_b^0=2.31$, which are shown in Table 1. 
In the following, we will extrapolate these values to the physical pion
mass with the formulas in Eq.(\ref{3ii}). 
 
\begin{table}
\caption{Extrapolated values of $a E_{\rm sim}$ at $a M_b^0=2.31$.}
\begin{center}
\begin{tabular}{lccc}
\hline
\hline
$\kappa$ &$a E_{\rm sim} (\Lambda_b)$  & $a E_{\rm sim} (\Sigma_b)$ 
&$a E_{\rm sim} (\Sigma_b^*)$   \\ 
\hline
0.13690 &0.816(33) &0.877(28) &0.889(27) \\
\hline
0.13750 &0.779(43) &0.845(32) &0.856(33) \\
\hline
0.13808 &0.733(63) &0.818(40) &0.827(37) \\
\hline
\hline
\end{tabular}
\end{center}
\end{table}
  
In our fit we have to determine three parameters in our formalism, 
($a_{\Sigma_b}$, $b_{\Sigma_b}$, and $\Lambda$ in Eq.(\ref{3ii}), 
for example). These parameters are related to $\Delta M$, $\alpha$, $g_1$, 
and $g_2$, which represent interactions at the quark and gluon level and 
cannot be determined from the chiral Lagragian for heavy baryons. 
In our fit, we treat them as effective parameters and assume that their 
possible slight $m_Q$ dependence, which results from QCD corrections
and $1/m_Q$ corrections, has been taken into account effectively
in this way. 

$\Delta M$ is the mass difference between sextet and antitriplet heavy baryons.
Since we do not have experimental data for the masses of $\Sigma_b^{(*)}$, we
use the data for $\Sigma_c^{(*)}$ to determine $\Delta M$ \cite{pdg}. 
The spin-averaged
mass of $\Sigma_c^{(*)}$ is $\frac{1}{6}(2m_{\Sigma_c}+4 m_{\Sigma_c^*})$,
which is bigger than $m_{\Lambda_c}$ by 0.213GeV. In our fit, we let 
$\Delta M$ vary between 0.17GeV and 0.23GeV, which are given by 
$m_{\Sigma_c}-m_{\Lambda_c}$ and $m_{\Sigma_c^*}-m_{\Lambda_c}$, respectively.
The mass difference $m_{\Sigma_c^*}-m_{\Sigma_c}$, which is equal to
$\frac{3\alpha}{m_c}$ to order $1/m_c$, leads to $\alpha=0.032$GeV$^2$ if
we choose $m_c=0.15$GeV. To see the depedence of our fit on $\alpha$, we
let it vary from 0.025GeV$^2$ to 0.035GeV$^2$.

The coupling constant $g_2$ can be determined from the decay width for
$\Sigma_c^* \ra \Lambda_c \pi$, which has the following explicit form:
\begin{equation}
\Gamma_{\Sigma_c^* \ra \Lambda_c \pi}=\frac{g_2^2}{12\pi f_\pi^2}\left[
\frac{(m_{\Sigma_c^*}^2-m_{\Lambda_c}^2)^2-2m_\pi^2(m_{\Sigma_c^*}^2
+m_{\Lambda_c}^2)+m_\pi^4}{4m_{\Sigma_c^*}^2}\right]^{\frac{3}{2}}
\frac{(m_{\Sigma_c^*}+m_{\Lambda_c})^2-m_\pi^2}{m_{\Sigma_c^*}^2}.
\label{4a}
\end{equation}
From $\Gamma_{\Sigma_c^{*++} \ra \Lambda_c \pi^+}=18\pm 5$GeV, we have
$g_2^2=0.559\pm 0.155$, while from $\Gamma_{\Sigma_c^{*0} \ra \Lambda_c \pi^-}
=13\pm 5$GeV, we have $g_2^2=0.404\pm 0.155$. Hence, in our fit we choose
the range $0.249 \le g_2^2 \le 0.714$.

Since $\Sigma_c^*$ cannot decay to $\Sigma_c \pi$, we cannot fix $g_1$ from
decays. However, $g_1$ can be related to the matrix of the axial-vector 
current between sextet heavy baryon states where a $u \ra d$ transition is
involved. By assuming $g_A^{ud}=0.75$, which corresponds to $g_A^{\rm nucleon}
=1.25$ in neutron $\beta$-decays and using spin-flavor wave functions for heavy
baryons, the authors in Ref.\cite{yan} found that $g_1=0.38$. Based on
this, we let $g_1^2$ vary from 0.1 to 0.2 in our fit. 

As discussed in Section III, the parameter $\Lambda$ characterizes the size
of the source of the pion. In principle, the value of $\Lambda$ can be
determined by fitting the lattice data.
However, since $\Lambda$ is mainly related to the data at small pion masses
and the current lattice data are only available at large pion masses, 
the error in the determination of $\Lambda$ is very large. 
The size difference between $\Sigma_b$ and $\Sigma_b^*$ is caused by 
effects of order $1/m_b$
which are small. The size difference between $\Sigma_b$ and 
$\Lambda_b$ is caused by the difference between $0^+$ and $1^+$ light degrees 
of freedom, which is also the main reason for a size difference between
$N$ and $\Delta$. It has been pointed out that the values of $\Lambda$ for 
$N$ and $\Delta$  are very close to each other \cite{lein1}. Hence we expect
that the difference between the values of $\Lambda$ for $\Sigma_b$ and 
$\Lambda_b$ should also be small. Since the integrand in $X_1$ becomes small
near the cutoff $\Lambda$, a small variation in $\Lambda$ will only lead to
an even smaller change in $X_1$. Based on these arguments, we will ignore the
differences among the values of $\Lambda$ for $\Sigma_b$, $\Sigma_b^*$,
and $\Lambda_b$. To see the dependence of our analysis on $\Lambda$, we
let $\Lambda$ vary between 0.4GeV and 0.6GeV.
 
Using the three masses for $\Sigma_b$, $\Sigma_b^*$,
and $\Lambda_b$ in Table 1, we fix the other two parameters (
$a_{\Sigma_b}$ and $b_{\Sigma_b}$ for $\Sigma_b$, for example) in 
Eq.(\ref{3ii}) with the least squares fitting method. The
values for these two parameters in the cases of $\Sigma_b^\pm$ and 
$\Sigma_b^{*\pm}$ are shown in Table 2, where we choose
$\Lambda=0.5$GeV, $\alpha=0.032$GeV$^2$, $\Delta M=0.213$GeV, $g_1^2=0.15$,
and $g_2^2=0.48$. The extrapolated masses for $\Sigma_b$, $\Sigma_b^*$,
and $\Lambda_b$ at the physical pion mass are also shown in this table.
The spin-averaged mass $m_{\Sigma_b}^{\rm ave}$ is defined as
$\frac{1}{6}(2m_{\Sigma_b}+4 m_{\Sigma_b^*})$.

\begin{table}
\caption{Fitted parameters, extrapolated masses of $\Sigma_b^\pm$, 
$\Sigma_b^{*\pm}$,
and $\Lambda_b$ and mass differences at $m_\pi^{\rm phys}$. Numbers in
brackets are errors caused by the errors of lattice data.}
\begin{center}
\begin{tabular}{cccc}
\hline
\hline
& $\Sigma_b^\pm$& $\Sigma_b^{*\pm}$& $\Lambda_b$ \\
\hline
$a ({\rm GeV})$ & 1.472(0.143)& 1.485(0.187)& 1.290(0.208)\\
\hline 
$b ({\rm GeV}^{-1})$ & 0.324(0.265)& 0.340(0.326)& 0.439(0.366)\\
\hline
$\bar{a} ({\rm GeV})$ & & 0.0188(0.0026)& \\
\hline 
$\bar{b} ({\rm GeV}^{-1})$ & & -0.00166(0.00470)& \\
\hline
$m ({\rm GeV})$ &1.4575(0.1384)& 1.4702(0.1803)& 1.2502(0.2008)\\
\hline
$m_{\Sigma_b^{*\pm}}-m_{\Sigma_b^\pm} ({\rm GeV})$ & &0.0127(0.2272)&  \\
\hline
$m_{\Sigma_b^\pm}^{\rm ave}-m_{\Lambda} ({\rm GeV})$& &0.2158(0.2385)&  \\
\hline
($m_{\Sigma_b^{*\pm}}-m_{\Sigma_b^\pm})^* ({\rm GeV})$ & &0.0180(0.0025)&  \\
\hline
\hline
\end{tabular}
\end{center}
\end{table}

With the parameters in Table 2 we obtain the masses of $\Sigma_b$, 
$\Sigma_b^*$, and $\Lambda_b$ as a function of the pion mass. These 
results are shown in Figs. 4, 5 and 6, respectively, for $\Lambda=0.4$ and
0.6. The difference between $m_{\Sigma_b}^{\rm ave}$ and $m_{\Lambda_b}$
is plotted in Fig. 7. The results of
linear extrapolation are also shown in these figures.

It can be seen from Table 2 that the extrapolated mass difference between 
$\Sigma_b$ and $\Sigma_b^*$ has a very large error. This is caused by the
uncertainty in the lattice data. A better way to obtain the mass difference 
between $\Sigma_b$ and $\Sigma_b^*$ is to extrapolate the lattice data
for this mass difference which were obtained from ratio fits, 
since these data have much smaller errors. The mass difference between 
$\Sigma_b$ and $\Sigma_b^*$, $\Delta E$, was also given in Ref.\cite{khan1}
for five different values of $a M^0$. We use the data at $a M^0$=
1.6, 2.0, 2.7, and 4.0 to obtain the value of $\Delta E$ at $a M^0$=2.31
with the formula 
\begin{equation}
a \Delta E=\frac{e}{a M^0},
\label{4b}
\end{equation}
where $e$ is a constant. Eq.(\ref{4b}) arises from the fact that the 
mass splitting between $\Sigma_b$ and $\Sigma_b^*$ is caused by
$1/m_Q$ effects. With the least squares fitting method we obtain
results for $\Delta E$ at $a M^0$=2.31, for different values of
$\kappa$. These are shown in Table 3.

\begin{table}
\caption{Extrapolated values of $\Delta E$, the mass difference between
$\Sigma_b^*$ and $\Sigma_b$, at $a M_b^0=2.31$ using Eq.(\ref{4b}).}
\begin{center}
\begin{tabular}{lccc}
\hline
\hline
$\kappa$ &0.13690  & 0.13750 &0.13808   \\ 
\hline
$a \Delta E$ &0.0093(5) &0.0095(7) &0.0096(7) \\
\hline
$\Delta E$ &0.0179(10) &0.0182(13) &0.0184(14) \\
\hline
\hline
\end{tabular}
\end{center}
\end{table}
 
In order to extrapolate the values in Table 3 to the physical mass of the
pion, we use the following formula:
\begin{equation}
m_{\Sigma_b^*}-m_{\Sigma_b}=\bar{a} + \bar{b} m_\pi^2 +\sigma_{\Sigma_b^*}-
\sigma_{\Sigma_b}.
\label{4c}
\end{equation}

With $\Lambda=0.5$GeV, $\alpha=0.032$GeV$^2$, $\Delta M=0.213$GeV, 
$g_1^2=0.15$, and $g_2^2=0.48$, we obtain $\bar{a}=0.0188(26)$, 
$\bar{b}=-0.00166(470)$, and the extrapolated mass diference between 
$\Sigma_b^\pm$ and $\Sigma_b^{*\pm}$ $m_{\Sigma_b^{*\pm}}-m_{\Sigma_b^\pm}
=0.0180(25)$,
which is listed in Table 2 as $(m_{\Sigma_b^{*\pm}}-m_{\Sigma_b^\pm})^*$. 
In Fig. 8, we show  $m_{\Sigma_b^{*\pm}}-m_{\Sigma_b^\pm}$ obtained in this way
as a 
function of the pion mass. For comparison, in Fig. 9 we plot the result for 
$m_{\Sigma_b^{*\pm}}-m_{\Sigma_b^\pm}$ which is obtained from Eq.(\ref{3ii}), 
although it has very large errors. Because these errors are so
large, the extrapolation from Eq.(\ref{3ii}) is consistent with the 
extrapolation 
directly from the lattice data for the mass difference between 
$\Sigma_b$ and $\Sigma_b^*$.

In addition to the uncertainties which are caused by 
the errors in the lattice data, the fitted results can also vary a little 
in the range of the parameters $\alpha$, $\Delta M$, $g_1^2$, $g_2^2$,
and $\Lambda$. 
In Table 4 we list these uncertainties.

\begin{table}
\caption{Uncertainties for the extrapolated quantities for $\Sigma_b^\pm$ and 
$\Sigma_b^{*\pm}$ which are caused by the
uncertainties associated with parameters in the fitting function.}
\begin{center}
\begin{tabular}{cccccc}
\hline
\hline
Quantities & $\alpha$ & $\Delta M$ & $g_1^2$ & $g_2^2$ &$\Lambda$ \\
\hline
$m_{\Sigma_b^0}$ &0.02\% & 0.3\% & 0.06\% & 0.8\% & 0.9\%\\
\hline
$m_{\Sigma_b^{*0}}$ &0.007\% & 0.3\% & 0.04\% & 0.9\% & 1.0\%\\
\hline
$m_{\Lambda_b}$ &0.0\% & 0.2\% & 0.0\% & 1.7\% & 1.8\%\\
\hline
$m_{\Sigma_b^{*0}}-m_{\Sigma_b^0}$ &3.1\% & 0.8\% & 1.6\% & 12.6\% & 10.2\%\\
\hline
$m_{\Sigma_b^0}^{\rm ave}-m_{\Lambda_b}$ &0.0\% & 3.4\% & 0.3\% & 3.7\% & 9.3\%\\
\hline
$(m_{\Sigma_b^{*0}}-m_{\Sigma_b^0})^*$ &2.2\% & 0.0\% & 1.1\% & 7.7\% & 1.7\%\\
\hline
\hline
\end{tabular}
\end{center}
\end{table}

In the naive linear extrapolations pion loop corrections are ignored. Hence
the results do not depend on the parameters $\alpha$, $\Delta M$, $g_1^2$, 
$g_2^2$, and $\Lambda$. In Table 5 we list 
the results of linear extrapolations for comparison. We note that
there is no difference between the results for $\Sigma_b^{(*)\pm}$
and $\Sigma_b^{(*)0}$ in the linear extrapolations.

\begin{table}
\caption{Fitted parameters, extrapolated masses of $\Sigma_b$, 
$\Sigma_b^{*}$, and $\Lambda_b$ and mass differences at 
$m_\pi^{\rm phys}$ for linear extrapolations. Numbers in
brackets are the errors caused by the errors of the lattice data.}
\begin{center}
\begin{tabular}{cccc}
\hline
\hline
& $\Sigma_b$& $\Sigma_b^{*}$& $\Lambda_b$ \\
\hline
$a ({\rm GeV})$ & 1.465(0.143)& 1.479(0.187)& 1.63(0.208)\\
\hline 
$b ({\rm GeV}^{-1})$ & 0.330(0.265)& 0.346(0.326)& 0.460(0.366)\\
\hline
$\bar{a} ({\rm GeV})$ & & 0.0190(0.0025)& \\
\hline 
$\bar{b} ({\rm GeV}^{-1})$ & & -0.00172(0.00470)& \\
\hline
$m ({\rm GeV})$ &1.4714(0.1384)& 1.4854(0.1803)& 1.2724(0.2008)\\
\hline
$m_{\Sigma_b^{*\pm}}-m_{\Sigma_b^\pm} ({\rm GeV})$ & &0.0140(0.2272)&  \\
\hline
$m_{\Sigma_b^\pm}^{\rm ave}-m_{\Lambda} ({\rm GeV})$& &0.2084(0.2385)&  \\
\hline
($m_{\Sigma_b^{*\pm}}-m_{\Sigma_b^\pm})^* ({\rm GeV})$ & &0.0190(0.0025)&  \\
\hline
\hline
\end{tabular}
\end{center}
\end{table}

Comparing the uncertainties listed in Table 2 and Table 4 we can see clearly
that the main uncertainties in our fit are caused by the errors in the lattice
data. In fact, the errors of lattice data for heavy baryons are much larger
than those for heavy mesons \cite{hein}. Indeed, the uncertainties
in the extrapolated heavy baryon masses are about one order larger than
those in the case of heavy mesons. However, because of the small errors in the
lattice data for the mass splitting between $\Sigma_b^\pm$ and 
$\Sigma_b^{*\pm}$ the extrapolated mass difference at the physical pion mass 
also has a smaller error, about 28\%. 

From Figures 4 to 9 we see that when the pion mass is smaller than about 
500MeV the extrapolations begin to deviate  from linear behavior.
This is because the pion loop corrections begin to affect the
extrapolations around this point. As the pion mass becomes smaller
and smaller, pion loop corrections become more and more important. 
For the masses of $\Sigma_b^\pm$, $\Sigma_b^{*\pm}$, $\Lambda_b$, and 
the mass difference between $\Sigma_b^\pm$ and $\Sigma_b^{*\pm}$ the 
extrapolated values are smaller than
those obtained by linear extrapolation. For the difference between the
spin-averaged mass of $\Sigma_b^{(*)\pm}$ and the mass of $\Lambda_b$,
the extrapolated value is larger than that obtained by linear 
extrapolation. We have checked that such behaviour is independent of 
the uncertainties in the parameters in our model.
 
Comparing the results in the naive linear extrapolations with those with 
pion loop corrections being included we find that the difference 
between them is much smaller than that found
in the case of mesons. For example, 
the splitting between $\Sigma_b^\pm$ and $\Sigma_b^{*\pm}$ is 
about 5\% smaller if pion loop effects are taken into account, while the
hyperfine splitting in the case of $B$ mesons is about 20\% smaller
when pion loop effects are taken into account \cite{guo}. For the masses
of $\Sigma_b^\pm$, $\Sigma_b^{*\pm}$, and $\Lambda_b$, the extrapolated
values with pion loop effects being included are about only 1\% smaller than
those in linear extrapolations, while for $B$ and $B^*$ the corresponding
number is about 3\% \cite{guo}.
Hence in the case of heavy baryons, the linear extrapolation is a better
approximation than in the case of heavy mesons. 

For $\Sigma_b^0$ and $\Sigma_b^{*0}$ we should use Eqs.(\ref{3ee}, \ref{3gg}) 
in the extrapolation of lattice data. Repeating the same procedure as that
for $\Sigma_b^\pm$ and $\Sigma_b^{*\pm}$ we find that, in addition to some
minor changes in numerical results, all the quantatitive results remain the 
same. In Tables 6 and 7 we list our numerical results for 
$\Sigma_b^0$ and $\Sigma_b^{*0}$. Comparing the results in Table 6 with those
in Tables 4 and 5 
we can see that the naive linear extrapolations work even better
in the case of $\Sigma_b^0$ and $\Sigma_b^{*0}$ than 
in the case of $\Sigma_b^\pm$ and $\Sigma_b^{*\pm}$.

\begin{table}
\caption{Fitted parameters, extrapolated masses of $\Sigma_b^0$ and 
$\Sigma_b^{*0}$ and mass difference at $m_\pi^{\rm phys}$. Numbers in
brackets are errors caused by the errors in the lattice data.}
\begin{center}
\begin{tabular}{ccc}
\hline
\hline
& $\Sigma_b^0$& $\Sigma_b^{*0}$ \\
\hline
$a ({\rm GeV})$ & 1.469(0.143)& 1.482(0.187)\\
\hline 
$b ({\rm GeV}^{-1})$ & 0.327(0.265)& 0.342(0.326)\\
\hline
$\bar{a} ({\rm GeV})$ & 0.0187(0.0026)& \\
\hline 
$\bar{b} ({\rm GeV}^{-1})$ & -0.00151(0.00470)& \\
\hline
$m ({\rm GeV})$ &1.4638(0.1384)& 1.4774(0.1803)\\
\hline
$m_{\Sigma_b^{*\pm}}-m_{\Sigma_b^\pm} ({\rm GeV})$  &0.0135(0.2272) & \\
\hline
$m_{\Sigma_b^\pm}^{\rm ave}-m_{\Lambda} ({\rm GeV})$& 0.2226(0.2385)& \\
\hline
($m_{\Sigma_b^{*\pm}}-m_{\Sigma_b^\pm})^* ({\rm GeV})$ & 0.0187(0.0025)&  \\
\hline
\hline
\end{tabular}
\end{center}
\end{table}

\begin{table}
\caption{Uncertainties for the extrapolated quantities for $\Sigma_b^0$ and 
$\Sigma_b^{*0}$ which are caused by the
uncertainties in the parameters of the fitting function.}
\begin{center}
\begin{tabular}{cccccc}
\hline
\hline
Quantities & $\alpha$ & $\Delta M$ & $g_1^2$ & $g_2^2$ &$\Lambda$ \\
\hline
$m_{\Sigma_b^\pm}$ &0.01\% & 0.2\% & 0.06\% & 0.4\% & 0.4\%\\
\hline
$m_{\Sigma_b^{*\pm}}$ &0.007\% & 0.1\% & 0.04\% & 0.5\% & 0.4\%\\
\hline
$m_{\Lambda_b}$ &0.0\% & 0.2\% & 0.0\% & 1.7\% & 1.8\%\\
\hline
$m_{\Sigma_b^{*\pm}}-m_{\Sigma_b^\pm}$ &1.5\% & 0.0\% & 2.2\% & 5.9\% & 1.5\%\\
\hline
$m_{\Sigma_b^\pm}^{\rm ave}-m_{\Lambda_b}$ &0.0\% & 2.3\% & 0.3\% & 6.6\% & 7.9\%\\
\hline
$(m_{\Sigma_b^{*\pm}}-m_{\Sigma_b^\pm})^*$ &1.6\% & 0.0\% & 1.1\% & 0.7\% & 0.5\%\\
\hline
\hline
\end{tabular}
\end{center}
\end{table}

\vspace{0.2in}
{\large\bf V. Summary and discussion}
\vspace{0.2in}

The masses of heavy baryons $\Sigma_b$, $\Sigma_b^{*}$, $\Lambda_b$,
and the mass difference between $\Sigma_b$ and $\Sigma_b^{*}$ have been
calculated numerically in lattice QCD with unphysical pion masses 
which are larger than about 560MeV. In order to extrapolate these data
to the physical mass of the pion in a consistent way, 
we included pion loop effects
on the heavy baryon masses by applying the effective chiral Lagrangian for 
heavy baryons when the pion mass is smaller than the inverse radii of heavy
baryons. This chiral Lagrangian is invariant under both
chiral symmetry (when the 
light quark masses go to zero) and heavy quark symmetry (when the heavy quark 
masses go to infinity). In order to study mass difference between $\Sigma_b$
and $\Sigma_b^{*}$, we took the color-magnetic-moment operator at order 
$1/m_Q$ in HQET into account since this operator is the leading one to
cause splitting between $\Sigma_b$ and $\Sigma_b^{*}$. When $m_\pi$ becomes 
large, lattice data show that heavy baryon masses depend on 
$m_\pi^2$ linearly in the range of interest. 
Based on these considerations, 
we proposed a phenomenological functional form to 
extrapolate the lattice data. 

The advantage of our formalism is that it
has the correct chiral limit behavior as well as the appropriate 
behavior when $m_\pi$ is large and that there are only three parameters
to be determined in the fit to lattice data.
It is found that when the pion mass is smaller than about 
500MeV the extrapolations begin to deviate from the naive linear 
extrapolations.
However, the differences between the extrapolations with and without
pion loop effects being included is smaller than those in the case of
heavy mesons. Hence the linear extrapolation is a better
approximation in the case of heavy baryons. We carefully
analysed uncertainties in our 
extrapolations which are caused by both lattice data errors and 
uncertainties in several
parameters in our model and found that the main uncertainties are 
caused by the errors of the lattice data. By directly extrapolating the lattice
data for $m_{\Sigma_b^*}-m_{\Sigma_b}$, which has much 
smaller errors, we found that the extrapolated
mass difference between $\Sigma_b^\pm$ and $\Sigma_b^{*\pm}$ at the
physical mass of the pion is 18.0 MeV, with an uncertainty of 28\% caused by 
lattice data errors. For $\Sigma_b^0$ and $\Sigma_b^{*0}$ this difference is
18.7MeV with 26\% uncertainty from lattice data errors. The uncertainties
associated with the parameters in our model are at most a few percent.
For the difference between the
spin-averaged mass of $\Sigma_b^{(*)\pm}$ and the mass of $\Lambda_b$,
the extrapolated value has a very large error. This needs to be improved
when the lattice data become more accurate.
Furthermore, we should bear in mind that our extrapolations are based on
the lattice data in the quenched approximation. From our experience in 
the cases of light and heavy mesons \cite{yoshie, guo}, the 
quenched approximation may affect the mass splitting
between $\Sigma_b$ and $\Sigma_b^{*}$. In addition, 
the lattice results for $m_{\Sigma_b^*}-m_{\Sigma_b}$ may be sensitive
to both the coefficient of the ${\bf \sigma\cdot B}$ term in NRQCD   
\cite{hein} and the clover coefficient in the clover action for light quarks.
This may also influence the lattice data and consequently affect
our extrapolations.

\vspace{2cm}

\noindent {\bf Acknowledgment}:

This work was supported by the Australian Research Council.

\newpage

\vspace{0.2in}

\noindent{\large \bf Figure Captions} \\
\vspace{0.4cm}

\noindent Fig. 1 Pion loop corrections to the propagator of $SU(3)$ 
sextet heavy baryons with spin-1/2, where $S_Q^{(*)}$ represent 
spin-1/2(3/2) $SU(3)$ 
sextet heavy baryons with heavy quark $Q$ and $T_Q$ represents $SU(3)$
antitriplet heavy baryons. 
\vspace{0.2cm}

\noindent Fig. 2 Pion loop corrections to the propagator of $SU(3)$ 
sextet heavy baryons with spin-3/2. Same notation as in Fig. 1.
\vspace{0.2cm}

\noindent Fig. 3 Pion loop corrections to the propagator of $SU(3)$ 
antitriplet heavy baryons. Same notation as in Fig. 1.
\vspace{0.2cm}

\noindent Fig. 4 Phenomenological fits to the lattice data for the masses
of $\Sigma_b^\pm$ as a function of the pion mass.  
The solid (dashed) line corresponds to $\Lambda=0.4$(0.6)GeV 
and the dotted line represents the fit using a linear extrapolation. 
\vspace{0.2cm}

\noindent Fig. 5 Phenomenological fits to the lattice data for the masses
of $\Sigma_b^{* \pm}$ 
as a function of the pion mass.  Same notation as in Fig. 4.
\vspace{0.2cm}

\noindent Fig. 6 Phenomenological fits to the lattice data for the masses
of $\Lambda_b$ as a function of the pion mass.  Same notation as in Fig. 4.
\vspace{0.2cm}

\noindent Fig. 7 Difference between the spin-averaged mass of 
$\Sigma_b^{(*)\pm}$ 
and the mass of $\Lambda_b$ as a function of the pion mass, which is obtained 
from Figs.4, 5 and 6 (same notation as in Fig. 4).
\vspace{0.2cm}

\noindent Fig. 8 Phenomenological fits to the lattice data for the mass
difference between $\Sigma_b^\pm$ and  $\Sigma_b^{*\pm}$ 
as a function of the pion mass  (same notation as in Fig. 4).
\vspace{0.2cm}

\noindent Fig. 9 $m_{\Sigma_b^{*\pm}}-m_{\Sigma_b^\pm}$ as a function of the 
pion mass, which is obtained from Figs.4 and 5. The large errors of lattice 
data are not shown (same notation as in Fig. 4).
\vspace{0.2cm}

\newpage

\begin{figure}[p]
\begin{center}
{\epsfsize=16in\epsfbox{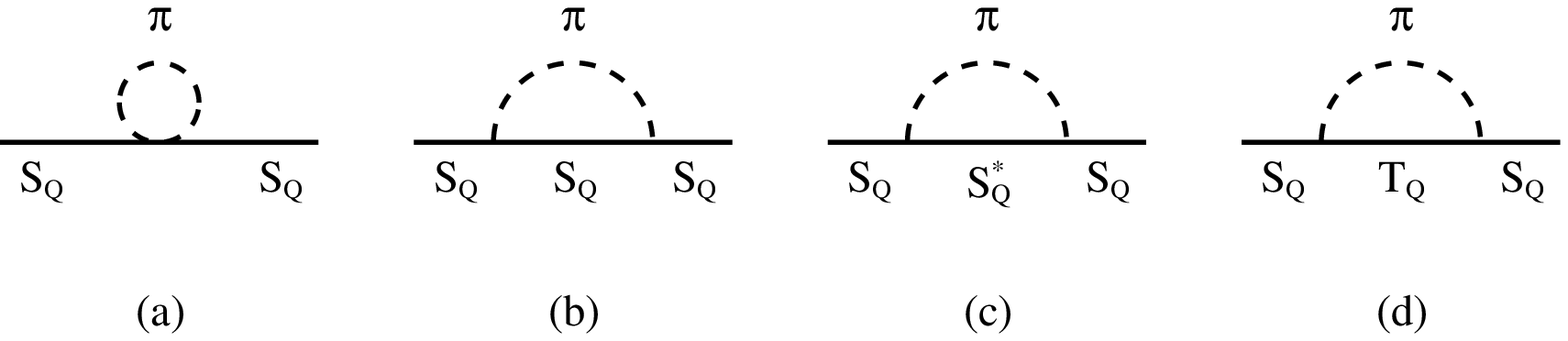}}
\end{center}
\vspace{0.cm}
\centerline{Fig. 1}
\end{figure}

\begin{figure}[p]
\begin{center}
{\epsfsize=16in\epsfbox{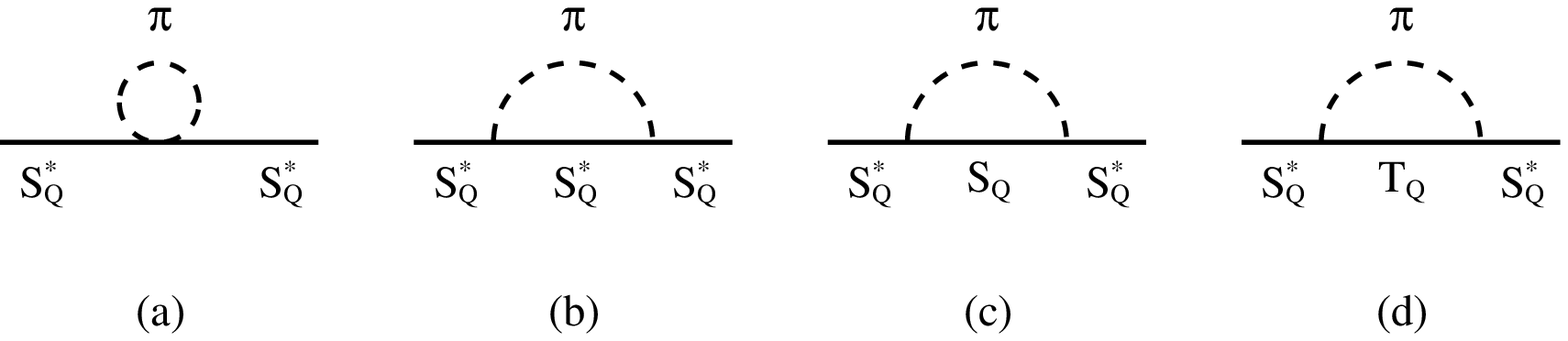}}
\end{center}
\vspace{0.cm}
\centerline{Fig. 2}
\end{figure}


\begin{figure}[p]
\begin{center}
{\epsfsize=15.1in\epsfbox{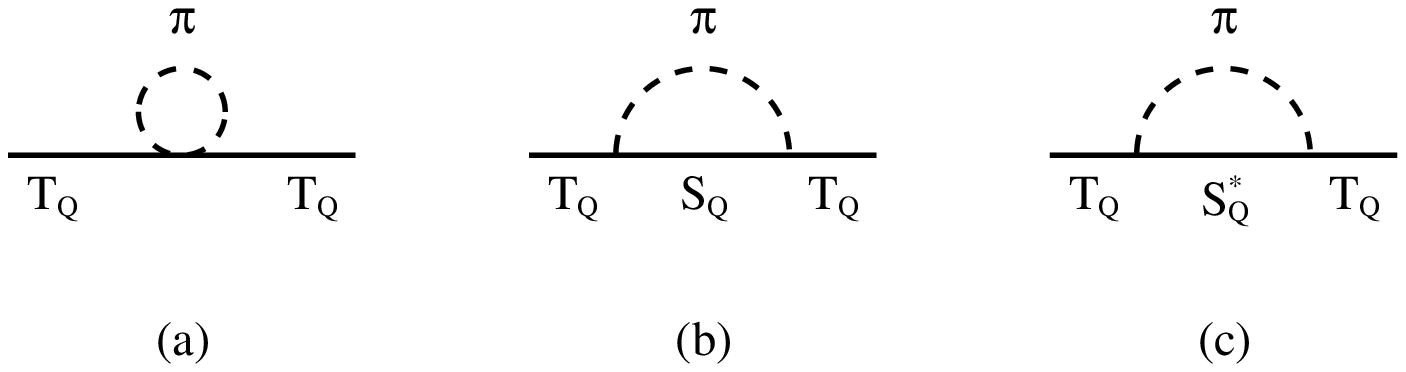}}
\end{center}
\vspace{0.cm}
\centerline{Fig. 3}
\end{figure}

\begin{figure}[p]
\begin{center}
{\epsfsize=14.5in\epsfbox{sigb.eps}}
\end{center}
\vspace{0.cm}
\centerline{Fig. 4}
\end{figure}

\begin{figure}[p]
\begin{center}
{\epsfsize=14.5in\epsfbox{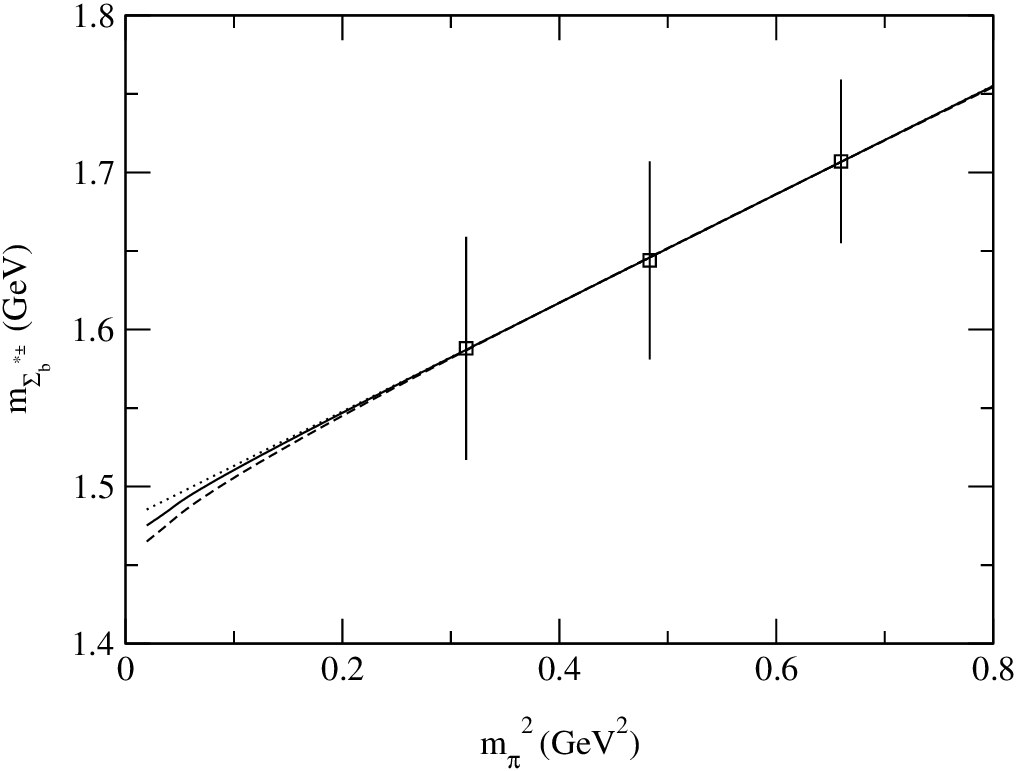}}
\end{center}
\vspace{0.cm}
\centerline{Fig. 5}
\end{figure}

\newpage

\begin{figure}[p]
\begin{center}
{\epsfsize=14.5in\epsfbox{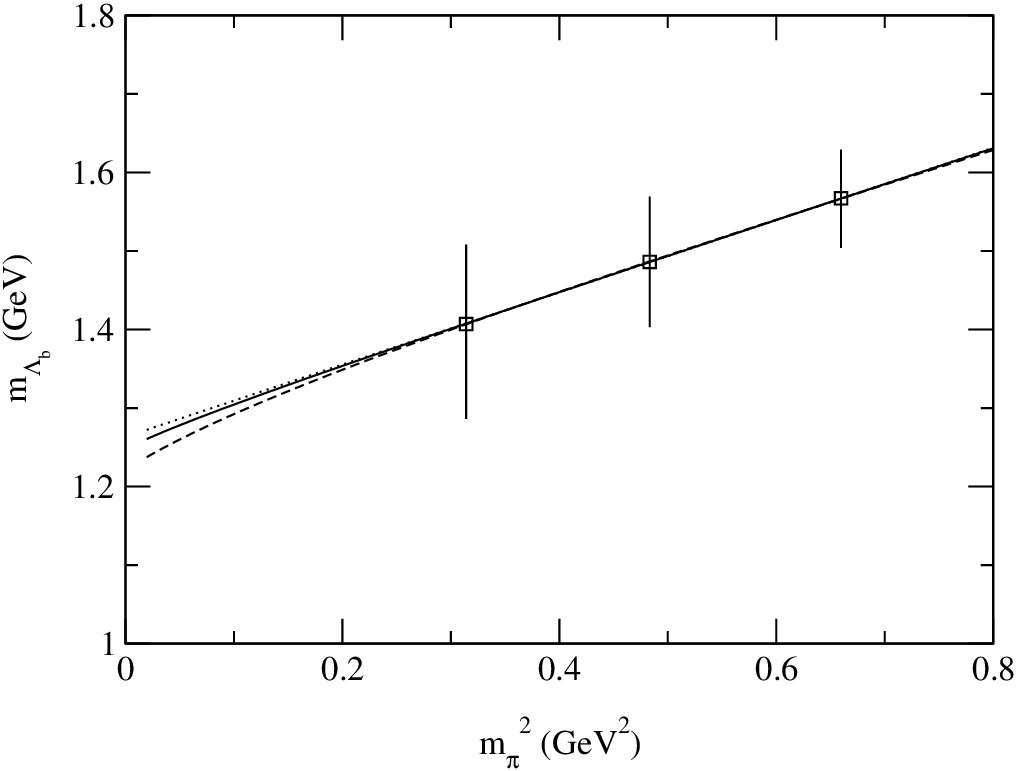}}
\end{center}
\vspace{0.cm}
\centerline{Fig. 6}
\end{figure}

\begin{figure}[p]
\begin{center}
{\epsfsize=14.5in\epsfbox{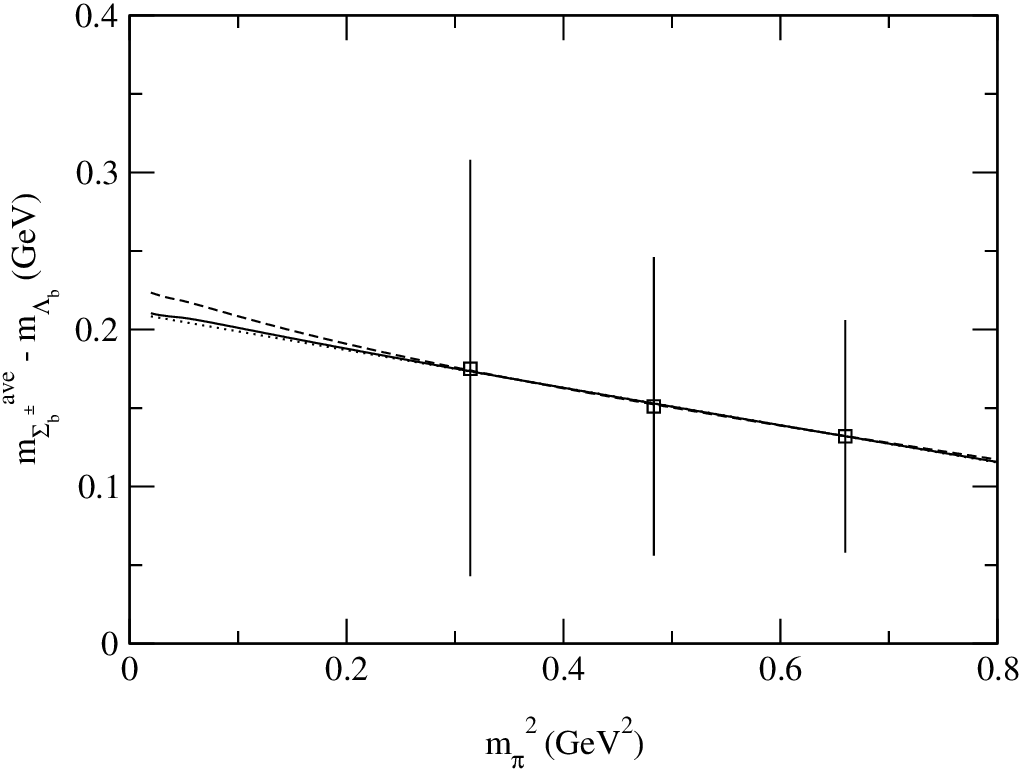}}
\end{center}
\vspace{0.cm}
\centerline{Fig. 7}
\end{figure}

\newpage

\begin{figure}[p]
\begin{center}
{\epsfsize=14.5in\epsfbox{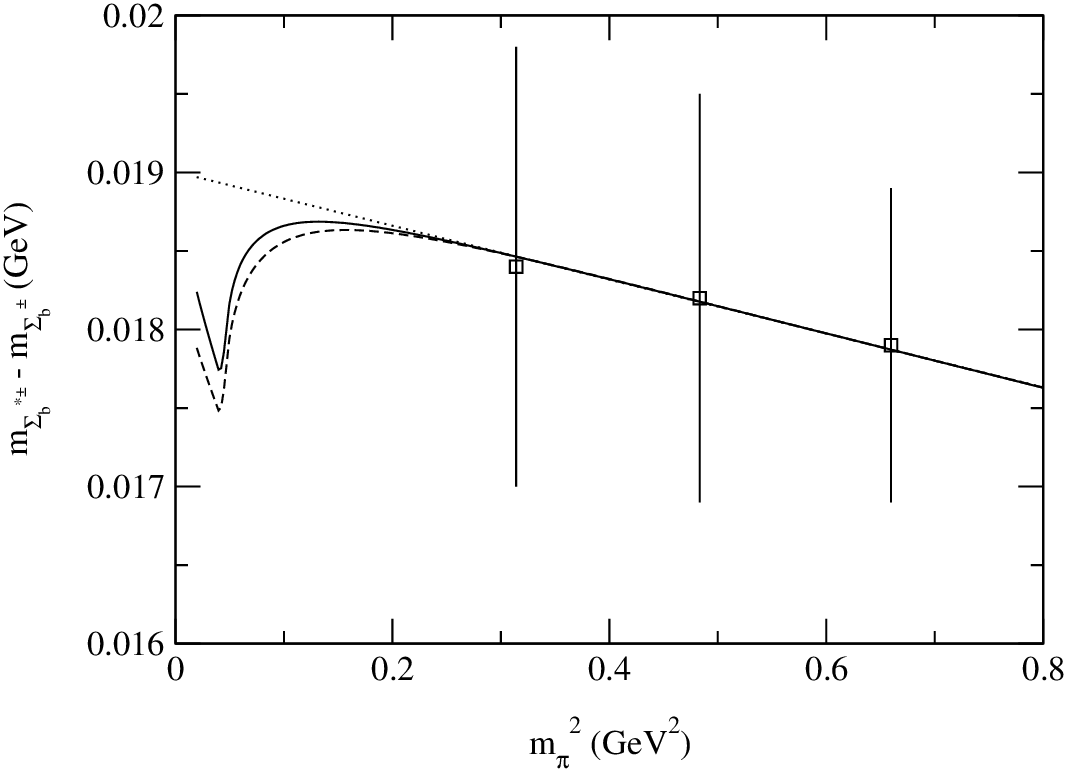}}
\end{center}
\vspace{0.cm}
\centerline{Fig. 8}
\end{figure}

\begin{figure}[p]
\begin{center}
{\epsfsize=14.5in\epsfbox{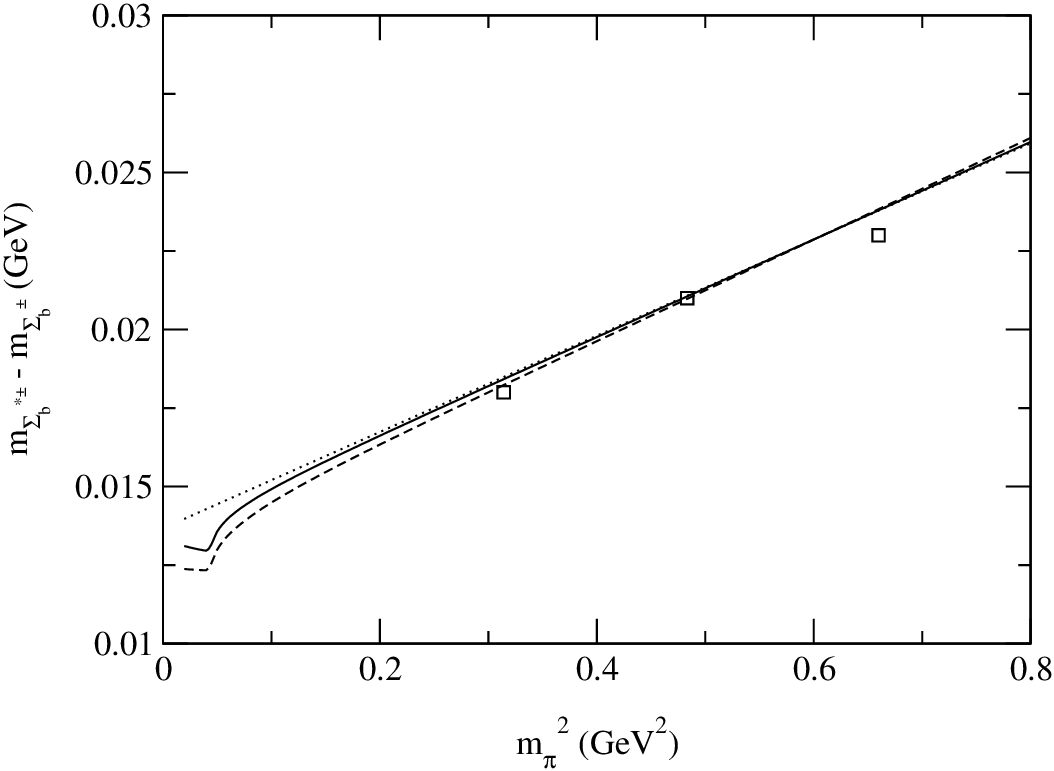}}
\end{center}
\vspace{0.cm}
\centerline{Fig. 9}
\end{figure}

\end{document}